\documentclass[final,times,5p]{elsarticle}
\pdfoutput=1
\usepackage{graphicx}
\usepackage{xspace}
\usepackage{amsmath}
\usepackage{amssymb}
\usepackage{amsfonts}
\usepackage{hyperref}

\usepackage{subfig}
\newcommand{\grad}{\ensuremath{{}^{\circ}}\xspace}
\newcommand{\ee}{\ensuremath{e^{+}e^{-}}\xspace}

\newcommand{\elel}{\ensuremath{\ell^{+}\ell^{-}}\xspace}
\newcommand{\ellell}{\ensuremath{\ell^{+}\ell^{-}}\xspace}

\newcommand{\mumu}{\ensuremath{{\mu^{+}\mu^{-}}}\xspace}

\newcommand{\Jpsi}{\ensuremath{J/\psi}\xspace}
\newcommand{\psip}{\ensuremath{\psi\text{(2S)}}\xspace}

\renewcommand{\Im}{\ensuremath{\text{Im}}\xspace}
\renewcommand{\Re}{\ensuremath{\text{Re}}\xspace}
\renewcommand{\epsilon}{\varepsilon}

\newcommand{\Gee}{\ensuremath{\Gamma_{ee}}\xspace}
\newcommand{\Gmumu}{\ensuremath{\Gamma_{\mu\mu}}\xspace}

\newcommand{\Gll}{\ensuremath{\Gamma_{\ell\ell}}\xspace}

\newcommand{\GBee}{\ensuremath{\Gamma_{ee}\times\mathcal{B}_{ee}}\xspace}
\newcommand{\GBmumu}{\ensuremath{\Gamma_{ee}\times\mathcal{B}_{\mu\mu}}\xspace}

\newcommand{\GBtautau}{\ensuremath{\Gamma_{ee}\times\mathcal{B}_{\tau\tau}}\xspace}
\newcommand{\GBhadr}{\ensuremath{\Gamma_{ee}\times\mathcal{B}_{\hadrons}}\xspace}

\newcommand{\Br}{\ensuremath{\mathcal{B}}\xspace}

\newcommand{\Bee}{\ensuremath{{\Br}_{ee}}\xspace}
\newcommand{\Bmumu}{\ensuremath{{\Br}_{\mu\mu}}\xspace}

\newcommand{\sigmaw}{\ensuremath{{\sigma}_{\scriptscriptstyle{W}}}\xspace}
\newcommand{\eff}{\ensuremath{\varepsilon}\xspace}

\newcommand{\deltasf}{\ensuremath{\delta_\text{sf}}\xspace}

\newcommand{\Scan}[1]{Scan #1}
\newcommand{\Point}[1]{Peak/cont. #1}

\newcommand{\scansh}[1]{sc. #1}
\newcommand{\pointsh}[1]{p/c #1}

\newcommand{\tof}{\text{tof}}
\newcommand{\theor}{\text{theor}}
\newcommand{\res}{\text{res}}
\newcommand{\cont}{\text{cont}}
\newcommand{\inter}{\text{int}}
\newcommand{\inters}{\text{s-int}}
\newcommand{\intert}{\text{t-int}}
\newcommand{\simul}{\text{sim}}
\newcommand{\observed}{\text{obs}}
\newcommand{\expected}{\text{expect}}
\newcommand{\eV}{\text{eV}\xspace}
\newcommand{\kev}{\text{keV}\xspace}
\newcommand{\MeV}{\text{MeV}\xspace}
\newcommand{\background}{\text{bg}}
\newcommand{\stat}{\text{stat}}
\newcommand{\syst}{\text{syst}}
\newcommand{\corr}{\text{corr}}

\newcommand{\nb}{\text{nb}}
\newcommand{\run}{\text{run}}

\newcommand{\ndf}{\text{n.\,d.\,f.}}

\newcommand{\hadrons}{\text{hadrons}}

\newcommand{\Result}{19.3}
\newcommand{\Stat}{0.3}
\newcommand{\Syst}{0.5}

\newcommand{\pdgGee}{2.34}
\newcommand{\pdgBmumu}{7.9}
\newcommand{\pdgeGee}{0.04}
\newcommand{\pdgeBmumu}{0.9}
\newcommand{\pdgResult}{18.5}
\newcommand{\pdgError}{2.1}

\newcommand{\eeResult}{21.2}
\newcommand{\eeStat}{0.7}
\newcommand{\eeSyst}{1.2}

\newcommand{\GeeLUResult}{2.279}
\newcommand{\GeeLUStat}{0.015}
\newcommand{\GeeLUSyst}{0.042}
\newcommand{\GeeResult}{2.282}
\newcommand{\GeeStat}{0.015}
\newcommand{\GeeSyst}{0.042}

\newcounter{syst}
\renewcommand{\sitem}[1]{\refstepcounter{syst}\thesyst \label{syst:#1} &}

\usepackage{color}
\newcommand{\showtableref}{}

\begin{document}
\hyphenation{ca-lo-ri-me-ter}

\begin{frontmatter}
	\title{Measurement of \GBmumu for \psip meson}

\author[binp]{V.V.~Anashin}
\author[binp]{O.V.~Anchugov}
\author[binp,nsu]{V.M.~Aulchenko}
\author[binp,nsu]{E.M.~Baldin}
\author[binp,nstu]{G.N.~Baranov}
\author[binp]{A.K.~Barladyan}
\author[binp,nsu]{A.Yu.~Barnyakov}
\author[binp,nsu]{M.Yu.~Barnyakov}
\author[binp,nsu]{S.E.~Baru}
\author[binp]{I.Yu.~Basok}
\author[binp]{A.M.~Batrakov}
\author[binp]{E.A.~Bekhtenev}
\author[binp,nsu]{A.E.~Blinov}
\author[binp,nsu,nstu]{V.E.~Blinov}
\author[binp,nsu]{A.V.~Bobrov}
\author[binp,nsu]{V.S.~Bobrovnikov}
\author[binp,nsu]{A.V.~Bogomyagkov}
\author[binp,nsu]{A.E.~Bondar}
\author[binp,nsu]{A.R.~Buzykaev}
\author[binp,nsu]{P.B.~Cheblakov}
\author[binp,nstu]{V.L.~Dorohov}
\author[binp,nsu]{S.I.~Eidelman}
\author[binp,nsu,nstu]{D.N.~Grigoriev}
\author[binp,nsu]{S.A.~Glukhov}
\author[binp]{S.E.~Karnaev}
\author[binp]{G.V.~Karpov}
\author[binp]{S.V.~Karpov}
\author[binp,nstu]{K.Yu.~Karukina}
\author[binp]{D.P.~Kashtankin}
\author[binp]{T.A.~Kharlamova}
\author[binp]{V.A.~Kiselev}
\author[binp]{V.V.~Kolmogorov}
\author[binp,nsu]{S.A.~Kononov}
\author[binp]{K.Yu.~Kotov}
\author[binp]{A.A.~Krasnov}
\author[binp,nsu]{E.A.~Kravchenko}
\author[binp,nsu]{V.N.~Kudryavtsev}
\author[binp,nsu]{V.F.~Kulikov}
\author[binp,nstu]{G.Ya.~Kurkin}
\author[binp]{I.A.~Kuyanov}
\author[binp,nsu]{E.A.~Kuper}
\author[binp,nstu]{E.B.~Levichev}
\author[binp,nsu]{D.A.~Maksimov}
\author[binp]{V.M.~Malyshev}
\author[binp,nsu]{A.L.~Maslennikov}
\author[binp,nsu]{O.I.~Meshkov}
\author[binp]{S.I.~Mishnev}
\author[binp]{I.A.~Morozov}
\author[binp,nsu]{I.I.~Morozov}
\author[binp,nsu]{N.Yu.~Muchnoi}
\author[binp]{S.A.~Nikitin}
\author[binp,nsu]{I.B.~Nikolaev}
\author[binp]{I.N.~Okunev}
\author[binp,nsu,nstu]{A.P.~Onuchin}
\author[binp]{S.B.~Oreshkin}
\author[binp,nsu]{A.A.~Osipov}
\author[binp,nstu]{I.V.~Ovtin}
\author[binp,nsu]{S.V.~Peleganchuk}
\author[binp,nstu]{S.G.~Pivovarov}
\author[binp]{P.A.~Piminov}
\author[binp]{V.V.~Petrov}
\author[binp,nsu]{V.G.~Prisekin}
\author[binp,nsu]{O.L.~Rezanova}
\author[binp,nsu]{A.A.~Ruban}
\author[binp]{G.A.~Savinov}
\author[binp,nsu]{A.G.~Shamov}
\author[binp]{D.N.~Shatilov}
\author[binp]{D.A.~Shvedov}
\author[binp,nsu]{B.A.~Shwartz}
\author[binp]{E.A.~Simonov}
\author[binp]{S.V.~Sinyatkin}
\author[binp]{A.N.~Skrinsky}
\author[binp,nsu]{A.V.~Sokolov}
\author[binp]{D.P.~Sukhanov}
\author[binp,nsu]{A.M.~Sukharev\corref{cor1}}
\ead{A.M.Suharev@inp.nsk.su}
\author[binp,nsu]{E.V.~Starostina}
\author[binp,nsu]{A.A.~Talyshev}
\author[binp,nsu]{V.A.~Tayursky}
\author[binp,nsu]{V.I.~Telnov}
\author[binp,nsu]{Yu.A.~Tikhonov}
\author[binp,nsu]{K.Yu.~Todyshev}
\author[binp]{A.G.~Tribendis}
\author[binp]{G.M.~Tumaikin}
\author[binp]{Yu.V.~Usov}
\author[binp]{A.I.~Vorobiov}
\author[binp,nsu]{V.N.~Zhilich}
\author[binp]{A.A.~Zhukov}
\author[binp,nsu]{V.V.~Zhulanov}
\author[binp,nsu]{A.N.~Zhuravlev}
 \address[binp]{Budker Institute of Nuclear Physics, 11, akademika
 Lavrentieva prospect,  Novosibirsk, 630090, Russia}
 \address[nsu]{Novosibirsk State University, 2, Pirogova street,  Novosibirsk, 630090, Russia}
 \address[nstu]{Novosibirsk State Technical University, 20, Karl Marx
  prospect,  Novosibirsk, 630092, Russia}
\cortext[cor1]{Corresponding author}

	\begin{abstract}
		The product of the electronic width of the \psip meson and the
		branching fraction of its decay to the muon pair was
		measured
		in the $\ee \to \psip \to \mumu$ process
		 using nine data sets corresponding to an integrated
                luminosity of about 6.5 pb$^{-1}$ collected with the KEDR detector
                at the VEPP-4M electron-positron collider:
		\[
			\GBmumu = \Result \pm \Stat \pm \Syst ~\eV.
		\]
		Adding the previous KEDR results on hadronic and leptonic channels,
		 the values of the
		\psip electronic width were obtained under two assumptions:
		either with the assumption of lepton
		universality
		\[
			\Gee = \GeeLUResult \pm \GeeLUStat \pm \GeeLUSyst ~\kev
		\] or without it, summing up hadronic and three independent leptonic channels
		\[
			\Gee = \GeeResult \pm \GeeStat \pm \GeeSyst ~\kev.
		\]

	\end{abstract}

\end{frontmatter}

\section{Introduction}\label{sec:intro}
The narrow charmonium states are frequently referred to as a hydrogen atom of
QCD.  Their electronic widths $\Gee$ are
rather well predicted by potential
models~\cite{Badalian:2008bi,lakhina-2006-74}, while the accuracy of the
QCD lattice calculations of $\Gee$ gradually approaches that
of the measurements~\cite{dudek-2006-73}.  The total and
leptonic widths of a hadronic resonance, $\Gamma$ and
$\Gll$, describe fundamental properties of the strong
potential~\cite{Brambilla2011}.

Besides, the value of the
electronic width of narrow charmonium resonances is required for
various
sum rules, e.g. for determination of the $c$--quark mass~\cite{cMass}.

An experimental study of the leptonic decays of a narrow charmonium
is important by itself and is also required for the determination of
its electronic and total widths.

In this paper we present a measurement of the product of the \psip
meson electronic width and its branching fraction to the \mumu pair,
\GBmumu. Such an experiment effectively means a measurement of the
area under the resonance curve of the \psip meson and requires data
taking at several center-of-mass (c.\,m.)
energy points or the precise knowledge of
the machine energy spread.
It is worth noting that the presentation of the result in
this form
is most suitable for fits performed by
the Particle Data Group~\cite{PDG-2016} while taking into account results
of different experiments.

A measurement of \GBee for the \psip meson is much
more difficult compared
 to \GBmumu due to a large background from the nonresonant
production of \ee pairs, unlike the \Jpsi case~\cite{KEDR:jpsi_leptonic_widths},
where the probabilities of leptonic decays
are about ten times bigger.
Another experimental difficulty relevant for both \GBee and \GBmumu
measurements is the presence of various \psip decay modes producing
the background which has to be explicitly taken into account.

\section{VEPP-4M collider and KEDR detector}\label{sec:VEPP}
The VEPP-4M collider~\cite{Anashin:1998sj} can operate in the broad
range of beam energies from 1 to 6 GeV. Its peak luminosity in the
\psip energy region is about~\(2\times10^{30}\,\text{cm}^{-2}\text{s}^{-1}\).

One of the main features of the VEPP-4M is the possibility of a precise
energy determination.
At VEPP-4M the relative accuracy of  energy calibration with the
resonant depolarization is about \(10^{-6}\).
Between calibrations the energy interpolation
in the \psip energy range has the accuracy of
\(6\cdot10^{-6}\) ($\sim10$~keV)~\cite{KEDR:final_psi_masses2015}.

To monitor the beam energy during data taking,
the infrared light Compton
backscattering technique~\cite{energy-okr-technique} is employed
 (with 50$\div$70~keV precision in the
charmonium region).

\begin{figure}[ht]
\includegraphics[width=\columnwidth]{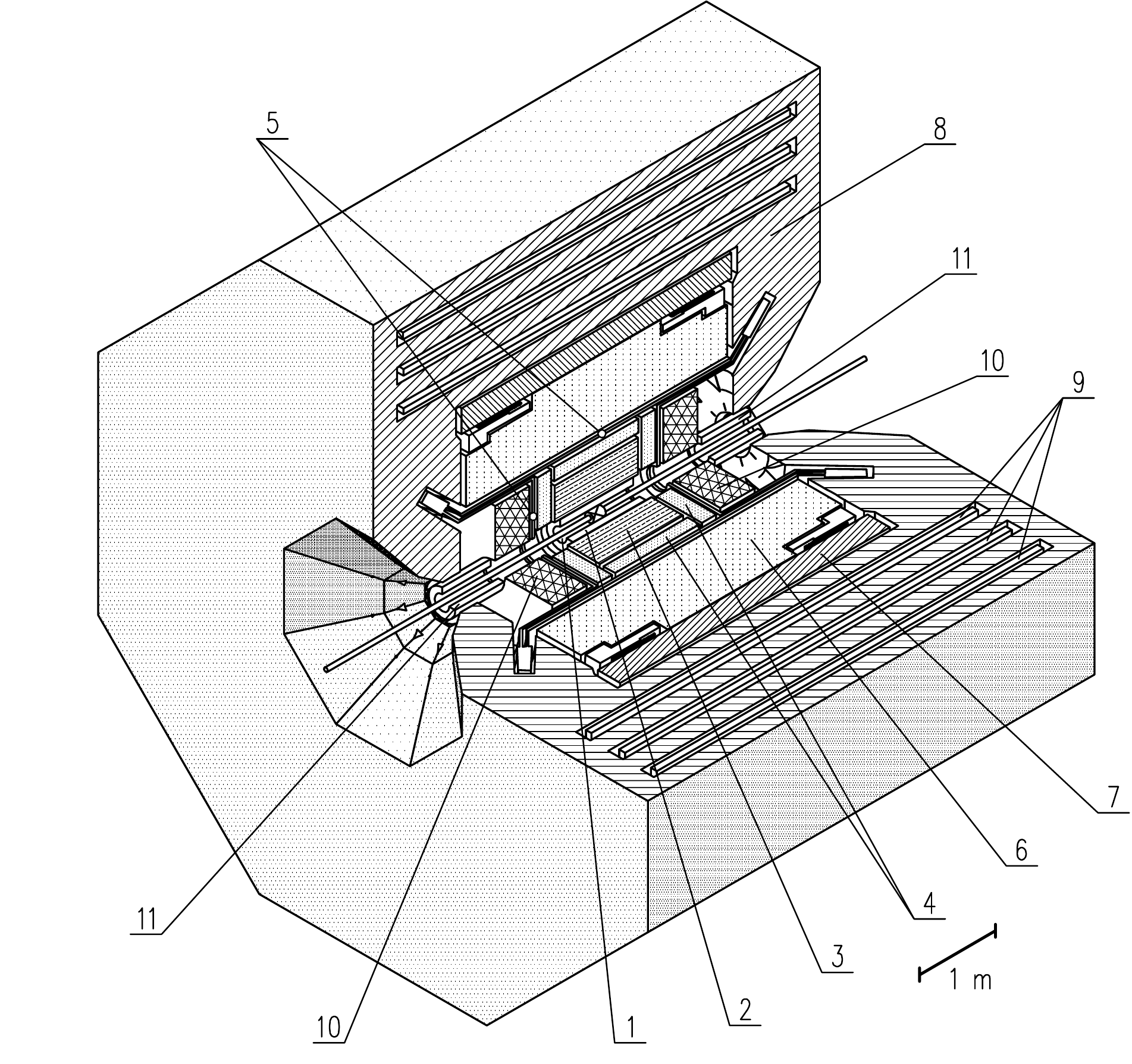}
\caption{The KEDR detector.
1~--~Vacuum chamber, 2~--~Vertex detector,
3~--~Drift chamber,
4~--~Threshold aerogel counters, 5~--~ToF counters,
6~--~ Liquid krypton calorimeter,
7~--~Superconducting solenoid, 
8~--~Magnet yoke, 9~--~Muon tubes, 10~--~CsI calorimeter,
11~--~Compensating superconducting coils.}
\label{kedr_picture}
\end{figure}

The main subsystems of the KEDR detector~\cite{KEDR2013} shown in
Fig.~\ref{kedr_picture} are the vertex detector, the drift chamber,
the scintillation time-of-flight (ToF) counters, the aerogel Cherenkov
counters, the barrel liquid krypton calorimeter, the endcap CsI calorimeter
and the three-layer muon system built in the yoke of a superconducting coil
generating a field of 0.65 T. The detector also includes a tagging system to
detect scattered electrons and study two-photon processes. The
on-line luminosity is counted by two independent
single-bremsstrahlung monitors.

\section{The experiment}\label{sec:experiment}
Several data sets in the \psip region were recorded with the KEDR detector
since 2004 (Table \ref{table:datasets}).

Two modes of data taking were employed. In the scan mode, the
experimental data were collected at several energy points around
the \psip resonance --- near the \psip cross section peak, at its
slopes, and in the continuum slightly below and above the resonance. In the
peak/continuum mode, only two energy points were recorded --- at the
peak and slightly below it. The exact positions of the energy points
varied with the data set.

\begin{table}[ht]\centering
\caption{KEDR \psip data sets}\label{table:datasets}
\medskip
\begin{tabular}{|l|l|c|c|}\hline
Data set & Period & \rule[-0.5em]{0cm}{1.5em} $\int L\,dt, \nb^{-1}$ & $\sigmaw$, \MeV \\\hline\hline
\Point{1} & January 2005 & 358  & 1.08 \\\hline
\Point{2} & Autumn 2005  & 222  & 0.99 \\\hline
\Scan{1}  & Spring 2006  & 255  & 0.99 \\\hline
\Point{3} & Spring 2006  & 631  & 0.99 \\\hline
\Point{4} & Autumn 2006  & 701  & 0.99 \\\hline
\Point{5} & Autumn 2007  & 1081 & 1.01 \\\hline
\Scan{2}  & End 2007  & 967  & 1.01 \\\hline
\Scan{3}  & Summer 2010   & 379  & 1.00 \\\hline
\Scan{4}  & End 2010  & 2005 & 0.98 \\\hline
\end{tabular}

\end{table}

A data sample considered in this analysis corresponds to a total integrated
luminosity of more than 6.5 pb$^{-1}$ or about $4\times 10^6$ \psip decays.

The collider energy spread \sigmaw required for cross section
determination was measured in scans using $\ee \to \hadrons$  with
accuracy of about 2\%~\cite{KEDR:final_psi_masses2015,KEDR:psi_masses2003,KEDR:main_parameters2012}.
The energy spread values measured in the most appropriate scans were
used for the peak/continuum data sets, yielding an additional
2\% uncertainty~\cite{KEDR:psi_masses2003} due to possible \sigmaw
variation between the data sets.

The detector conditions (trigger parameters, status of subsystems, etc.)
varied significantly during the experiment.
Various data sets could
be considered as partially independent.

The primary trigger signal was provided by a coincidence of two
non-adjacent ToF scintillation counters or a localized energy
deposition in the barrel calorimeter, for which the hardware energy
threshold of approximately 400 MeV has the width about 20\%.
 A veto
from the endcap-calorimeter crystals closest to the beam line was used
to suppress the machine background in data sets up to the
``peak/continuum~4'' data set inclusively.

The secondary trigger conditions were suitable to accept two-prong \ee
and \mumu events and changed in data sets according to the actual
detector state.

During the offline analysis both
 real and simulated events pass through
the software event filter (a so-called ``software trigger'') which
recalculates a
trigger decision using a digitized response of the detector subsystems.
This allows to exclude the uncertainties and instabilities of
the hardware thresholds.

\section{Theoretical \texorpdfstring{$\ee\to\ellell$}{electron-positrion
 to lepton-antilepton} cross section}\label{sec:theory}
The analytical expressions for the cross section of
the process $\ee\to\ellell$
with radiative corrections taken into account in the soft-photon
approximation were first derived by  Ya.\,A.~Azimov et al.
in 1975~\cite{azimov-1975-eng}.
With some up-today modifications one obtains
in the vicinity of a narrow resonance

\newcommand{\diffcs}{\left(\frac{d\sigma}{d\Omega}\right)}

\begin{equation}
  \label{eq:ee2mumu}
  \begin{aligned}
    &\diffcs^{ee\to\mu\mu} =
    \diffcs_{\text{QED}}^{ee\to\mu\mu}+
    \frac{3}{4W^2}
    \left(1+\delta_{\text{sf}}\right)
    \left(1+\cos^2\theta\right)
    \,\times\\
    &\quad\qquad\left\{
        \frac{3\Gamma_{ee}\Gamma_{\mu\mu}}{\Gamma M}
        \Im f-
        \frac{2\alpha\sqrt{\Gamma_{ee}\Gamma_{\mu\mu}}}{M}\,
        \Re \frac{f}{1-\Pi_0}
     \right\} ,
  \end{aligned}
\end{equation}
where a correction $\deltasf$ follows from the
structure function approach
of~\cite{KuraevFadinEng}:
\begin{equation}\label{eq:deltasf}
  \deltasf=\frac{3}{4}\beta+
   \frac{\alpha}{\pi}\left(\frac{\pi^2}{3}-\frac{1}{2}\right)+
  \beta^2\left(\frac{37}{96}-\frac{\pi^2}{12}-
  \frac{1}{36}\ln\frac{W}{m_e} \right)
\end{equation}
and
\begin{equation}  \label{eq:F}
  f=\frac{\pi\beta}{\sin\pi\beta}\,
      \left(\frac{M/2}{-W+M-i \Gamma/2}\right)^{1-\beta}
\end{equation}
with
\begin{equation}
\label{eq:beta}
\beta=\frac{4\alpha}{\pi}\left(\ln\frac{W}{m_e}-\frac{1}{2}\right).
\end{equation}
Here $W$ is the center-of-mass energy, $M$, $\Gamma$, $\Gee$ and $\Gmumu$
are the resonance mass and its total, electron and muon widths, respectively,
$\theta$
is the polar angle of outgoing particles, $\alpha$ is the fine-structure constant,
$m_e$ is the electron mass.
$\Pi_0$ represents the
vacuum polarization operator with the resonance
contribution excluded. The terms proportional to $\Im f$ and
$\Re f$ describe the contribution of the resonance
and the interference effect, respectively, and, using $\Bmumu=\Gmumu/\Gamma$,
could be rewritten as
\begin{equation}\begin{aligned}
\diffcs^{\mu\mu}_\res &\propto \frac{\GBmumu}{M}
\Im f \left(1+\cos^2\theta\right), \\
\diffcs^{\mu\mu}_\inter &\propto \frac{2\alpha\sqrt{\Gamma\,\GBmumu}
}{M} \Re f\frac{1}{1-\Pi_0} \left(1+\cos^2\theta\right),
\end{aligned}\end{equation}
with clearly shown parameter \GBmumu and common angular dependence.

The leptonic width definition in
Eqs.~\eqref{eq:ee2mumu}--~\eqref{eq:beta}
implicitly  includes vacuum polarization
as  recommended by PDG:
$\Gamma_{\ell\ell}=\Gamma^{0}_{\ell\ell}/|1-\Pi_0|^2$, where $\Gamma^{0}_{\ell\ell}$
is the lowest-order QED value.

In the \ee channel one has
\begin{equation}
  \label{eq:ee2ee}
  \begin{aligned}
    &\diffcs^{ee\to ee} =
    \diffcs_{\text{QED}}^{ee\to ee}+\\
    &\frac{\left(1+\deltasf\right)}{W^2}
    \left\{\frac{9}{4}\frac{\Gamma^2_{ee}}{\Gamma
      M}(1+\cos^2\theta) \,\Im f \right.-\\
    &\left.\frac{3\alpha}{2}\frac{\Gamma_{ee}}{M}
    \left [\frac{1+\cos^2\theta}{1-\Pi_0(s)}-
      \frac{(1+\cos\theta)^2}{(1-\cos\theta)}\frac{1}{1-\Pi_0(t)}\right ]
              \Re f
    \right\},\\
&s = W^2, t \approx -\frac{s}{2}(1-\cos\theta),
  \end{aligned}
\end{equation}
which leads to resonance and interference cross sections expressed
with $\Bee = \Gee/\Gamma$ as

\begin{equation}\begin{aligned}
\diffcs^{ee}_{\res~\,} &\propto \frac{\GBee}{M}
\Im f \left(1+\cos^2\theta\right), \\
\diffcs^{ee}_{\inters} &\propto \frac{2\alpha\sqrt{\Gamma\,\GBee}
}{M} \Re f \left(1+\cos^2\theta\right)\frac{1}{1-\Pi_0(s)},\\
\diffcs^{ee}_{\intert} &\propto \frac{2\alpha\sqrt{\Gamma\,\GBee}
}{M} \Re f
\frac{\left(1+\cos\theta\right)^2}{\left(1-\cos\theta\right)}\frac{1}{1-\Pi_0(t)},
\end{aligned}\end{equation}
where $\inters$ and $\intert$ parts correspond to s- and t-channel interference, respectively.
The $\res$ and $\inters$ parts have the same angular distribution while
$\intert$ has another one.

The accuracy of Eq.~\eqref{eq:ee2mumu} and \eqref{eq:ee2ee}
is about 0.1\%. Recently it was verified in the work~\cite{Zhou:2017yod}
where more precise analytical expressions of the cross sections were
suggested.

To compare experimental data with the theoretical cross
sections,
it is necessary to perform their convolution with a distribution of the total
beam energy which is
assumed to be Gaussian with an energy spread \(\sigmaw\):
\[\rho(W)=\frac{1}{\sqrt{2\pi}\,\sigmaw}\exp{\left(-\frac{(W-W_0)^2}{2\sigmaw^2}\right)}\,,\]
where \(W_0\) is an average c.m. collision energy.
Possible deviation of the distribution from the Gaussian was studied
in the analysis~\cite{KEDR:final_psi_masses2015}, and it is taken
into account as a systematic uncertainty.

\section{Data analysis}\label{sec:data}

\subsection{Event selection}
Events satisfying the criteria below were selected as \mumu:
\begin{enumerate}
 \item There should be exactly two oppositely charged tracks
 originating from the beam collision point. Each track should have a
 corresponding energy deposit in the calorimeter.
 \item The software trigger gives a positive decision.
 \item Polar $\theta$ and asimutal $\varphi$ acollinearity is $<28\grad$.
 \item The energy deposited in the calorimeter for each track should
 not exceed 700 \MeV.
 \item There should be not more than one cluster in the calorimeter
 which is not associated with the tracks, and its energy should not
 exceed 160 \MeV.
 \item Both tracks are confirmed by the muon system. Namely, a
 ``confirmed'' track should have at least one associated hit in the
 first or second layer of the muon system.
 \item Both tracks should be in the polar angle range of the muon
 system: $50\grad<\theta<130\grad$.\label{cuts:mumu:theta_lim}
\end{enumerate}

Additionally, there was a time-of-flight condition
to suppress cosmic background. The
condition applied to experimental data only. Its efficiency was
measured explicitly
and discussed in detail
in section~\ref{sec:tofeff}.

\ee events were selected using the following requirements:
\begin{enumerate}
 \item There should be exactly two oppositely charged tracks
 originating from the beam collision point. Each track should have a
 corresponding energy deposit in the calorimeter.
 \item The software trigger gives a positive decision.
 \item Polar $\theta$ and asimutal $\varphi$ acollinearity is $<28\grad$.
 \item The energy deposited in the calorimeter for each track should be
 greater than 800 \MeV.
 \item There should be not more than one cluster in the calorimeter
 which is not associated with the tracks, and its energy should not
 exceed 160 \MeV.
 \item Both tracks should be in the polar range of the barrel
 calorimeter: $45\grad<\theta<135\grad$.
\end{enumerate}

No ToF condition is applied to \ee events.

\subsection{Fit procedure}\label{subsec:fit}
 The luminosity was measured using Bhabha
scattering with the resonance contribution properly taken into account.
In the peak/continuum
data sets this implies to use the difference of angular distributions of
Bhabha and resonance events.
To extract the luminosity, \ee events were
divided into equal angular intervals (four by default) of the polar
angle $\theta$ from $45\grad$ to
$135\grad$. An ''average'' $\theta = (\pi - \theta_+ + \theta_-) / 2$
was used for each event to take into account possible detector asymmetries.
 At the $i$-th energy \(W_i\) and the
$j$-th angular interval \(\theta_j\), the expected number of
$\ee\to\ee$ events was parameterized as

\begin{equation}\label{eq:number_of_ee}\begin{aligned}
N^\expected_{\ee}(W_i,\theta_j)=&\mathcal{L}_i\cdot\sigma^\expected_{ee}(W_i, \theta_j),\\
\sigma^\expected_{ee}(W_i, \theta_j) =&
(\sigma^\theor_\res(W_i,\theta_j) + \sigma^\theor_{\inters}(W_i,\theta_j))\cdot
\eff_\res(\theta_j)|_i\\
+&\sigma^\theor_{\intert}(W_i,\theta_j)\cdot
\eff_\inter(\theta_j)|_i\\
+&\sigma^\simul_\cont(W_i,\theta_j)\cdot
\eff_\cont(\theta_j)|_i\\
+&\sigma^\expected_\background(W_i, \theta_j),
\end{aligned}\end{equation}
where $\mathcal{L}_i$ --- integrated luminosity at $W_i$,
$\sigma^\theor$ --- theoretical cross sections for elastic scattering,
resonance and interference, $\eff(\theta_j)|_i$ --- detection
efficiencies for the $j$-th angular interval obtained from simulation. The
last sum element is the expected contribution of background processes.
Each contribution has its own angular distribution and thus its own
detection efficiency. The interference angular distribution consists
of two parts, one of them being the same as resonance and another with
separate $\eff_\intert$.

Since there are no angular $\theta$ bins for \mumu, the expected number
of events at the energy $W_i$ is just:
\begin{equation}\label{eq:number_of_mumu}\begin{aligned}
N^\expected_\mumu(W_i)=& \mathcal{L}_i\cdot\sigma^\expected_{\mu\mu}(W_i),\\
\sigma^\expected_{\mu\mu}(W_i)=& \left.\eff_\tof^\observed\right|_i\times
\Big((\sigma^\theor_\res(W_i) +
\sigma^\theor_\inter(W_i))\cdot\eff_\res|_i+\\
+&\sigma^\simul_\cont\cdot\eff_\cont|_i + \sigma^\expected_\background(W_i)
\Big),
\end{aligned}\end{equation}
which also includes the measured ToF efficiency
$\eff_\tof^\observed$. The resonance and interference for muons have
equal angular distributions and thus equal efficiencies.

The products of continuum cross sections and the detection efficiencies
$\sigma^\simul_\cont\cdot\eff_\cont$ for both \ee and \mumu are
calculated with the simulation program which also accounts for the
radiative corrections.

The expected background contribution is a sum of the background decay modes:
\begin{equation}
\sigma^\expected_\background (W_i)= \sum_m \sigma^\theor_{m}(W_i) \eff_{m}|_i,
\end{equation}
where $\eff_{m}$ --- mode $m$ efficiency (individually for each
$\theta$ bin in the \ee case), and its theoretical cross section
$\sigma^\theor_{m}(W)$ is calculated using the mode branching ratio
$\Br_{m}$. Various accelerator
and cosmic backgrounds were negligible and therefore were not included
in the background contribution.

The products $\GBmumu$ and $\GBee$ are free parameters of the fit.

\subsection{Simulation}\label{subsec:simulation}
For simulating the nonresonant contribution $\sigma_\cont$
in case of \ee
scattering
we use the
BHWIDE~\cite{BHWIDE} generator, MCGPJ~\cite{MCGPJ} and BABAYAGA~\cite{BABAYAGA}
being the
alternatives. The main generator for $\mumu$ scattering was MCGPJ.

The resonant and interference cross sections were simulated using
simple generators with proper angular distributions.
In this case the initial-state radiative corrections are already
taken into account in the expressions~\eqref{eq:ee2mumu} and \eqref{eq:ee2ee}.
These formulae implicitly involve the branching ratios
$\Gamma_{\ell\ell}/\Gamma = \mathcal{B}_{\ell\ell(n\gamma)}$
with the arbitrary number of soft photons emitted. Actual event
selection criteria can not
be 100\% efficient for events with additional photons,
therefore the final-state radiation
must be simulated explicitly. This was done using the PHOTOS
package~\cite{photos}.

The accelerator and cosmic backgrounds
as well as various detector noises might overlap useful events,
modifying their signature.
To take this effect into account,
the random trigger (RND) events were recorded during
the experiment, and, at the simulation processing stage, simulated
events were superimposed with the RND events.

\begin{table}[ht]\centering
\caption{\psip decay background (Bg) modes accounted for in the \mumu
analysis. Efficiencies and resulting corrections from each mode
vary with the data
sets.}\label{table:sim_bg}
\begin{tabular}{|l|c|c|c|}\hline
Bg mode & $\Br_m$, \% & Efficiency, \%
& Correction, \% \\ \hline \hline
$\Jpsi \pi^+ \pi^-$ & $34.49$ & $ 0.03 \div 0.09 $& $2.29\div8.94$\\\hline
$\Jpsi \pi^0 \pi^0$ & $18.16$ & $ 0.01 \div 0.02 $& $0.38\div0.92$\\\hline
$\gamma \chi_{c0}(1P)$ & $9.99$& $ < 0.01 $& $0.00\div0.05$\\\hline
$\gamma \chi_{c1}(1P)$ & $9.55$ & $ 0.03 \div 0.03 $& $0.47\div0.92$\\\hline
$\gamma \chi_{c2}(1P)$ & $9.11$ & $ 0.02 \div 0.03 $& $0.44\div0.69$\\\hline
$\Jpsi \eta$ & $3.36$ & $ 0.02 \div 0.05 $& $0.17\div0.46$\\\hline
$\ee$ & $0.79$& $ < 0.01 $& $ < 0.01 $\\\hline
$\eta_c \gamma$ & $0.34$& $ < 0.01 $& $ < 0.01 $\\\hline
$\tau^+\tau^-$ & $0.31$ & $ 0.05 \div 0.08 $& $0.05\div0.07$\\\hline
$\Jpsi \pi^0$ & $0.13$ & $ 0.10 \div 0.15 $& $0.03\div0.05$\\\hline
$p \bar{p}$ & $0.03$ & $ 0.01 \div 0.03 $& $ < 0.01 $\\\hline
\end{tabular}
\end{table}

Many \psip decays with \mumu or \ee in the final states could emulate
the effect events. For instance, in case of the cascade decay $\psip \to
\Jpsi \text{X} \to \elel \text{X}$, when X is undetected or
not correctly reconstructed,
the similarity could be complete. To subtract such a
contribution, simulation has to be used.

Table \ref{table:sim_bg} lists the background decay modes accounted
for. These modes have the largest branching ratios. The efficiencies
vary notably with the data sets due to significant changes of the
detector conditions, the most important one being the turn off
several layers of the drift chamber.

Multihadronic and two-photon processes were simulated as well~\cite{KEDR:generators_tayursky}. The
corresponding contributions were found to be negligible.

\section{Time-of-Flight measurement efficiency}\label{sec:tofeff}
Due to dead time in digitization electronics, the time-of-flight
measurement has a
significant inefficiency of about 10\%.
The trigger signals from the time-of-flight system
are routed through the separate channels and thus
are not affected by this inefficiency.

\begin{figure}[ht!]
\centering
\subfloat[$\ee\to$ $\mumu$
signal with cosmic
background]{\label{fig:t1_vs_t2:mumu}\includegraphics[width=0.45\columnwidth]{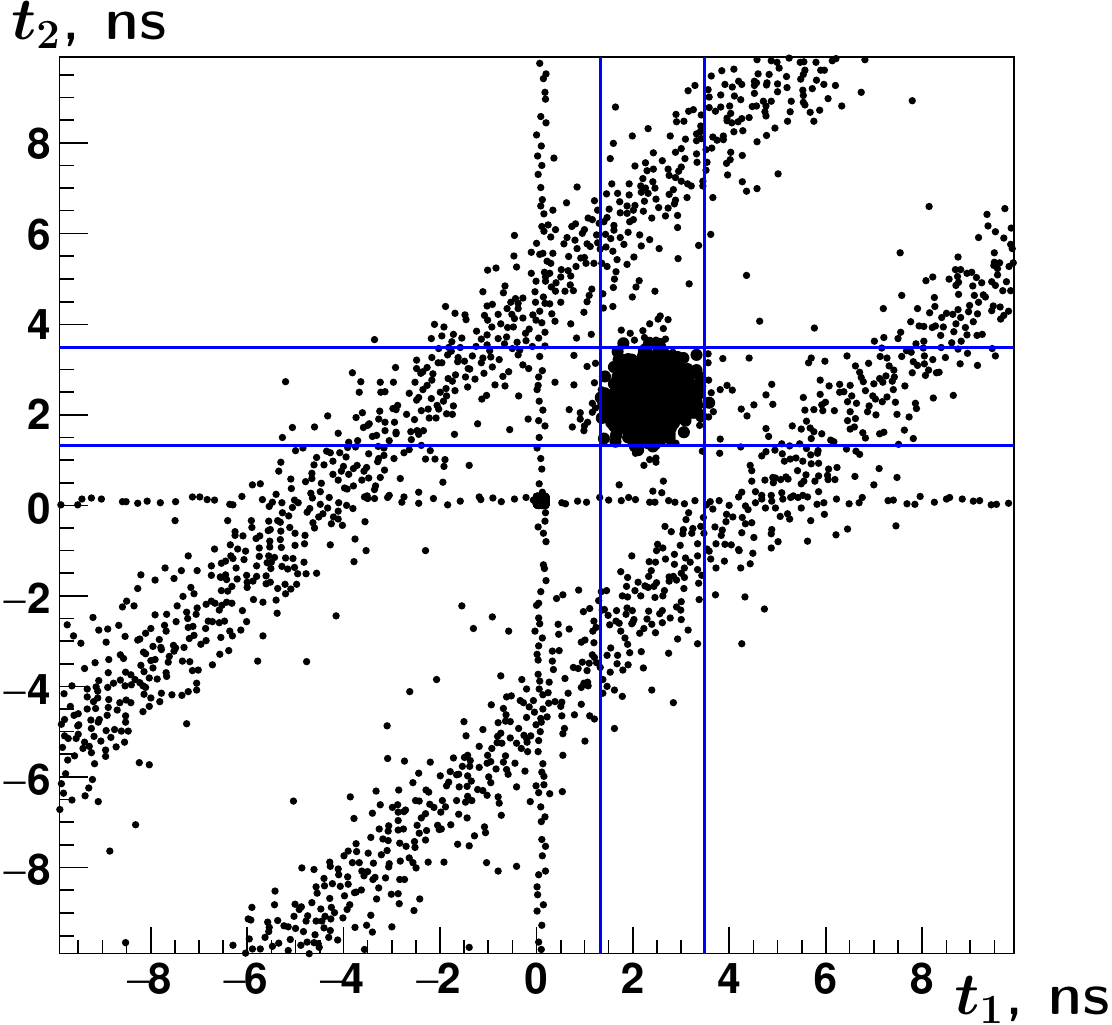}}\hfill\subfloat[cascade
decay to \mumu through $\Jpsi\pi\pi$]{\label{fig:t1_vs_t2:cascade}\includegraphics[width=0.45\columnwidth]{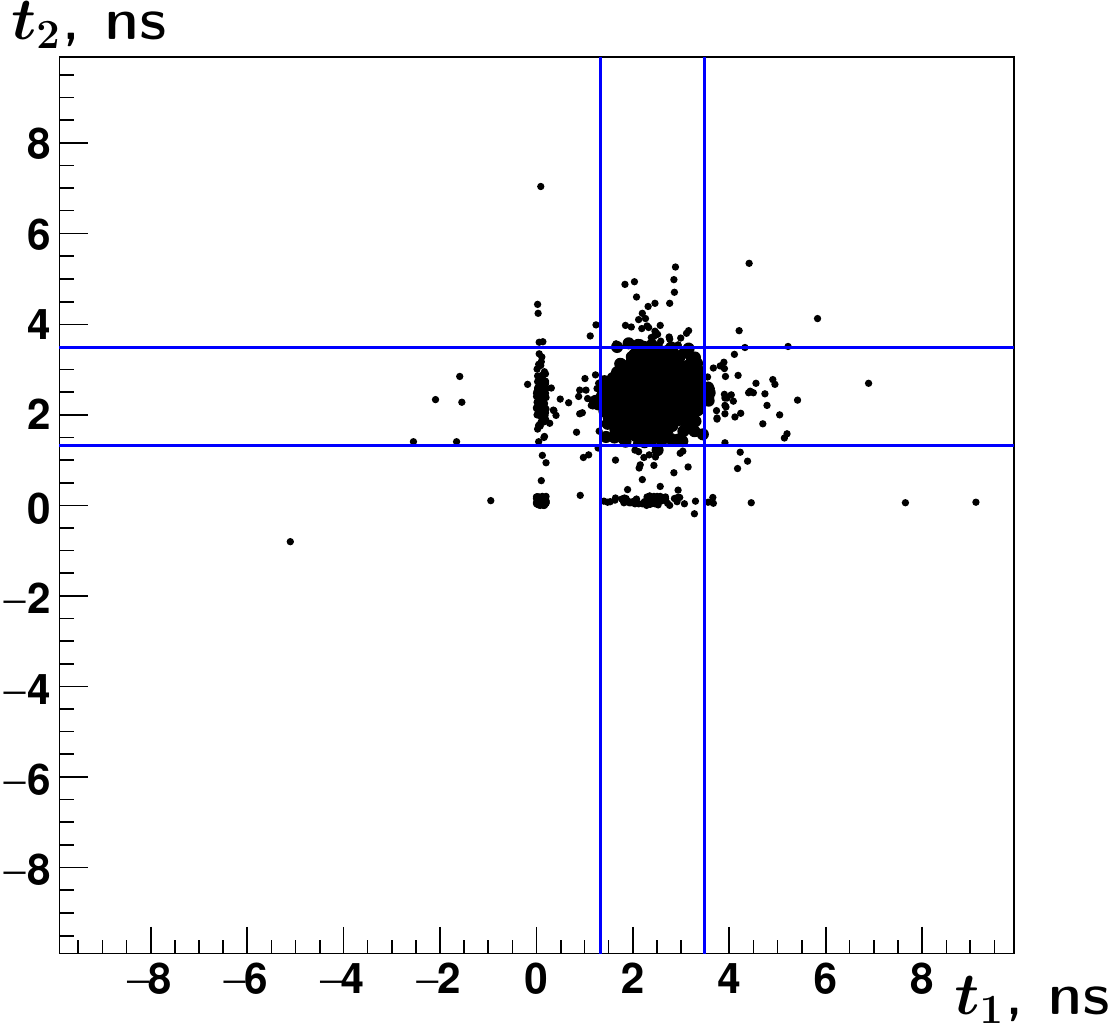}}
\caption{Time-of-flight distribution example for \mumu candidates. $t_{1,2}$
are the times of flight for muon candidate tracks, irrespective to their charge.
The
selection criteria are shown with a square. Events with lost time are located at
$t_{1,2} = 0$. Slant stripes correspond to cosmic events.}\label{fig:t1_vs_t2}
\end{figure}

The condition for each of selected tracks is:
\begin{equation}
| t \times \sin \theta - T_0| \leqslant 3\sigma_\tof,
\end{equation}
where $t$ and $\theta$ --- time of flight and polar angle, $T_0 = 2.4$
ns --- time of flight for \psip decay muons in the detector transverse plane,
 $\sigma_\tof = 0.36$ ns
--- the time resolution.
Fig.~\ref{fig:t1_vs_t2:mumu} shows a two-dimensional distribution of
the time of flight and the selection criteria.

\begin{figure}[ht]\centering
\includegraphics[width=\columnwidth]{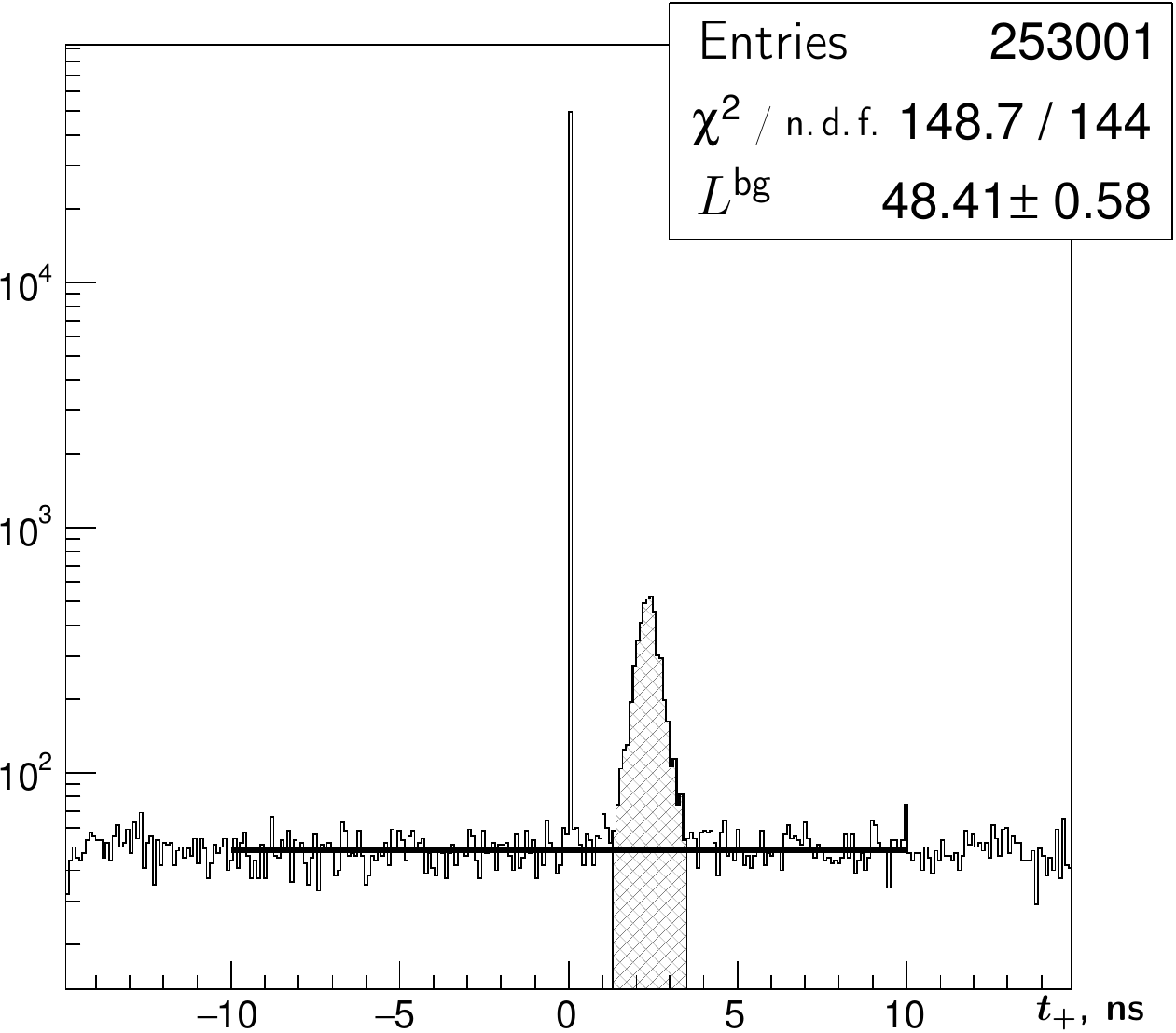}
\caption{An example of the time of flight ($t_+$) distribution
 for $\mu^+$ candidates without the $\mu^-$
condition
applied. The fitted cosmic background level $L^{bg}$ is shown with
the horizontal line, events passing the $\mu^+$
condition are selected by the dashed area.}\label{fig:tofeff_1}
\end{figure}

The efficiency measurement method (also applied in similar
KEDR analysis of \Jpsi data~\cite{KEDR:lepton_universality_2014}) estimates
the efficiencies for $\mu^+$ and $\mu^-$ separately:
\begin{equation}
\eff_{+/-} = \frac{N^\text{full}}{N_{-/+} - N^\text{bg}_{-/+}},\;
N^\text{bg}_i = 2\frac{3\sigma_\tof}{\Delta t} L^\text{bg}_i,
\end{equation}
where $N_{+/-}$ --- the number of events passing the condition for a
corresponding track, $N^\text{full}$
--- that passing the conditions for both tracks. The number of cosmic background
events is estimated from the time-of-flight distributions
(Fig.~\ref{fig:tofeff_1}), where
$L^\text{bg}_i$ is the fitted flat background level, $\Delta t$ --- bin width.

The total efficiency $\eff_\tof = \eff_+\eff_-$. The uncertainty due to
possible correlation of $\eff_+$ and $\eff_-$ is less than 0.3\%~\cite{KEDR:lepton_universality_2014}.

The results are presented in Table~\ref{table:tofeff_table}. Systematic
uncertainties for $\eff_\tof$ were obtained varying
conditions of the cosmic background fit
and accounting for a possible $\eff_+/\eff_-$ correlation. The
values of $\eff_\tof$ were additionally cross-checked using muons from
the cascade processes $\psip \to
\Jpsi\pi^+\pi^-, \psip \to \Jpsi\pi^0\pi^0$, where $\Jpsi$ decays into $\mumu$
with  times of flight very similar to muons from direct \psip decay
(Fig.~\ref{fig:t1_vs_t2:cascade}).

\begin{table}[ht]\centering
\caption{Time-of-flight selection criteria efficiency
for each data set
with statistical and systematic
errors.}\label{table:tofeff_table}
\begin{tabular}{|p{0.4\columnwidth}|c|}\hline
Data set & \parbox{0.4\columnwidth}{\centering$\eff_\tof$, \%} \\\hline\hline
\Point{1} & $ 85.9 \pm 0.7  \pm 0.9$  \\ \hline
\Point{2} & $ 83.6 \pm 1.0  \pm 1.1$  \\ \hline
\Scan{1} & $ 84.2 \pm 1.0  \pm 0.5$  \\ \hline
\Point{3} & $ 81.5 \pm 0.6  \pm 0.5$  \\ \hline
\Point{4} & $ 79.8 \pm 0.5  \pm 0.4$  \\ \hline
\Point{5} & $ 86.7 \pm 0.4  \pm 0.4$  \\ \hline
\Scan{2} & $ 82.9 \pm 0.5  \pm 1.1$  \\ \hline
\Scan{3} & $ 80.4 \pm 0.8  \pm 0.9$  \\ \hline
\Scan{4} & $ 81.7 \pm 0.4  \pm 0.9$  \\ \hline

\end{tabular}
\end{table}

\section{Systematic uncertainties}\label{sec:syst}
\begin{table*}[th!]
\newcommand{\systableheader}{Systematic uncertainty source}
\newcommand{\systablesigmaw}{C. m. energy distribution}
\newcommand{\systableMGtable}{Fixed values of $M_{\psip}$, $\Gamma_{\psip}$}
\newcommand{\systableE}{Energy measurement}
\newcommand{\systablePDG}{$\psip$ decay parameters}
\newcommand{\systablesimbhabha}{Bhabha simulation}
\newcommand{\systablesimmumu}{\mumu scattering simulation}
\newcommand{\systablecoll}{Collinearity cuts}
\newcommand{\systablethetamin}{\ee polar angle range}
\newcommand{\systablecharge}{Charge determination}
\newcommand{\systableEo}{Extra energy deposit cut}
\newcommand{\systablemu}{Muon system cut}
\newcommand{\systableabgnoise}{ABG noise}
\newcommand{\systableabgthr}{ABG thresholds}
\newcommand{\systabletrgthr}{Calo trigger thresholds}
\newcommand{\systablernd}{RND trigger application}
\newcommand{\systablephotos}{FSR accounting}
\newcommand{\systablefit}{\ee events $\theta$ binning}
\newcommand{\systabletofeff}{ToF measurement efficiency}
\newcommand{\systablePTeff}{Trigger efficiency}
\newcommand{\systabletheory}{Theoretical accuracy}
\newcommand{\systablefooter}{Sum in quadrature}
\newcommand{\systableasymmetry}{Detector asymmetry}

\centering\caption{Main sources of systematic uncertainties and their
relative contributions, \%.}\label{table:syst_table}
\begin{tabular}{| c|l | c|c|c|c|c|c|c|c|c|c|}\hline
\multicolumn{2}{|c|}{\systableheader} & \pointsh{1} & \pointsh{2} & \scansh{1} & \pointsh{3} & \pointsh{4} & \pointsh{5} & \scansh{2} & \scansh{3} & \scansh{4}& $\sigma_{\syst}^{\corr}$\\\hline\hline
\sitem{sigmaw} \systablesigmaw & $1.9$ & $2.7$ & $1.1$ & $2.9$ & $2.2$ & $2.6$ & $1.1$ & $2.9$ & $1.7$ & $0$\\\hline
\sitem{MGtable} \systableMGtable & $0.7$ & $0.6$ & $0.1$ & $0.3$ & $0.7$ & $0.7$ & $0.5$ & $0.2$ & $0.9$ & $0.1$\\\hline
\sitem{E} \systableE & $3.1$ & $0.6$ & $<0.1$ & $1.7$ & $0.3$ & $0.5$ & $0.2$ & $3.8$ & $2.7$ & $<0.1$\\\hline
\sitem{simbhabha} \systablesimbhabha & $1.4$ & $1.4$ & $2.2$ & $1.7$ & $1.1$ & $2.1$ & $1.6$ & $2.6$ & $0.9$ & $0.9$\\\hline
\sitem{simmumu} \systablesimmumu & $0.2$ & $0.2$ & $0.3$ & $0.2$ & $0.2$ & $0.2$ & $0.2$ & $0.3$ & $0.3$ & $0.2$\\\hline
\sitem{coll} \systablecoll & $0.8$ & $2.8$ & $2.4$ & $0.8$ & $2.1$ & $1.4$ & $1.5$ & $5.4$ & $1.6$ & $0.8$\\\hline
\sitem{thetamin} \systablethetamin & $1.1$ & $2.0$ & $1.8$ & $1.0$ & $1.0$ & $1.2$ & $1.6$ & $2.1$ & $1.3$ & $1.0$\\\hline
\sitem{charge} \systablecharge & $0.6$ & $0.3$ & $0.8$ & $0.6$ & $0.2$ & $1.9$ & $0.1$ & $1.0$ & $0.4$ & $0.1$\\\hline
\sitem{asymmetry} \systableasymmetry & $0.9$ & $0.2$ & $0.5$ & $0.9$ & $0.1$ & $0.1$ & $0.2$ & $0.4$ & $0.2$ & $0.1$\\\hline
\sitem{Eo} \systableEo & $1.4$ & $1.2$ & $2.2$ & $0.5$ & $1.0$ & $0.6$ & $2.2$ & $1.7$ & $1.6$ & $0.5$\\\hline
\sitem{mu} \systablemu & $2.5$ & $2.7$ & $2.2$ & $0.6$ & $0.3$ & $0.5$ & $0.6$ & $0.7$ & $<0.1$ & $0$\\\hline
\sitem{abgthr} \systableabgthr & $0.3$ & $0.7$ & $0.5$ & $0.1$ & $0.3$ &  ---  &  ---  &  ---  &  ---  & $0.1$\\\hline
\sitem{trgthr} \systabletrgthr & $0.1$ & $0.1$ & $0.2$ & $0.1$ & $<0.1$ & $0.4$ & $0.5$ & $0.4$ & $0.2$ & $<0.1$\\\hline
\sitem{rnd} \systablernd & $0.2$ & $0.1$ & $<0.1$ & $<0.1$ & $<0.1$ & $0.3$ & $0.1$ & $0.9$ & $0.3$ & $<0.1$\\\hline
\sitem{photos} \systablephotos & $0.4$ & $0.4$ & $0.4$ & $0.4$ & $0.4$ & $0.4$ & $0.4$ & $0.4$ & $0.3$ & $0.3$\\\hline
\sitem{fit} \systablefit & $0.6$ & $0.2$ & $0.6$ & $0.5$ & $0.5$ & $0.3$ & $0.1$ & $0.4$ & $0.3$ & $0.1$\\\hline
\sitem{tofeff} \systabletofeff & $1.9$ & $2.5$ & $1.5$ & $1.2$ & $0.8$ & $0.9$ & $2.8$ & $2.7$ & $2.3$ & $0.8$\\\hline
\sitem{PTeff} \systablePTeff & $0.9$ & $<0.1$ & $0.2$ & $0.1$ & $0.1$ & $0.1$ & $0.2$ & $0.1$ & $0.1$ & $<0.1$\\\hline
\sitem{theory} \systabletheory & $0.1$ & $0.1$ & $0.1$ & $0.1$ & $0.1$ & $0.1$ & $0.1$ & $0.1$ & $0.1$ & $0.1$\\\hline
\hline
\multicolumn{2}{|c|}{\systablefooter} & $5.7$ & $6.2$ & $5.4$ & $4.4$ & $3.7$ & $4.5$ & $4.7$ & $8.7$ & $4.9$ & $1.9$\\\hline
\end{tabular}

\end{table*}

The data sets used in the analysis are considered as
semi-independent experiments with independent statistical errors but
with partially correlated systematic errors. To obtain the final
result, the following weighting
procedure is applied:

\begin{equation}\begin{aligned}
\langle \GBmumu \rangle &= \sum w_i \times (\GBmumu)_i,\\
\sigma^2_{\stat} &= \sum w_i^2 \times \sigma^2_{\stat,i}\\
\sigma^2_{\syst} &= \sum w_i^2 \times (\sigma^2_{\syst,i}-\sigma^2_{\syst,0}) + \sigma^2_{\syst,0} \\
w_i &\propto 1/(\sigma^2_{\stat,i} + \sigma^2_{\syst,i} - \sigma^2_{\syst,0}),
\end{aligned}\end{equation}
where $w_i$ --- $i$-th measurement weight, accounting for statistical
errors and uncorrelated part of systematic errors,
$\sigma^2_{\syst,0}$ --- correlated part of systematic uncertainty.

The determination of the correlated part of systematic errors is not a
trivial task, and in most cases we assumed that the correlated part
corresponds to the minimal uncertainty in data sets for a given
uncertainty source. This leads to the most conservative estimates of the
total uncertainty.

Table~\ref{table:syst_table} shows the main sources of systematic
uncertainties, their contributions for each data set, correlated parts
and sums in quadrature. Below is the explanation for each row of
the table.

The contribution of the center-of-mass energy shape (\showtableref row
\ref{syst:sigmaw}) was obtained varying \sigmaw according to its
uncertainty, which
is partially due to non-Gaussian effects (see
sections~\ref{sec:experiment} and \ref{sec:theory}).

To calculate cross sections, the \psip mass value measured by the KEDR
detector~\cite{KEDR:final_psi_masses2015} was used, while the width was
taken from the PDG tables~\cite{PDG-2016}. Varying the mass and width
within their errors gives the contribution to the uncertainty of the result
(\showtableref row \ref{syst:MGtable}).

To estimate the uncertainty originated from the energy measurement
(\showtableref row \ref{syst:E}), the default analysis version
using average $\langle W_\run\rangle|_i$ to compute the cross section
$\sigma_i = \sigma(\langle W_\run\rangle|_i)$ at the $i$-th energy
point was compared to that with an average cross section $\sigma_i
= \langle\sigma(W_\run)\rangle|_i$.

The systematic uncertainty from \ee-scattering simulation was estimated
comparing results obtained with the default generator BHWIDE
and alternative generators BABAYAGA and MCGPJ.
The \mumu continuum cross section is calculated by the MCGPJ generator with
statistical precision of $\approx 0.1\%$, the systematic part $\lesssim
0.2\/$\% is the declared precision of the MCGPJ generator.
Estimations of these systematic uncertainties are presented in the
\showtableref
rows
\ref{syst:simbhabha} and \ref{syst:simmumu}.

The ``collinearity cut'' contribution to the resulting uncertainty (\showtableref
 row \ref{syst:coll}) was estimated by varying
the cuts: reducing the acollinearity limit from 28\grad to 10\grad and
imposing global polar angle limits just on one
track instead of both.

A minimal polar angle limit for \ee events varied in the interval
$[-5\grad, +5\grad]$ from its default value of 45\grad. Maximal change
of the result with such variations is shown in the \showtableref row
\ref{syst:thetamin}.

The charge misidentification impact (\showtableref row
\ref{syst:charge}) was studied using a special version of track
reconstruction allowing incorrect determination of one or two track
charges.

In the standard fit version, the ``average'' $\theta$
 angle (see subsection \ref{subsec:fit}) was used to build distributions
 over polar angle.
The ``detector asymmetry'' (\showtableref row
\ref{syst:asymmetry}) error comes from comparison of results obtained
with angles of positively $(\theta_+)$ or negatively $(\theta_-)$
charged tracks separately.

The requirement on calorimeter clusters unattached to any track was varied
widely, allowing two extra clusters instead of one, and limiting the
extra energy at 90 and 200 MeV instead of default 160 MeV. The maximal
change of the result is presented in the \showtableref row \ref{syst:Eo}.

The muon system cut contribution (\showtableref row
\ref{syst:mu}) was checked allowing one track without muon system
confirmation. To avoid extra background, an additional cut on particle
momenta has to be imposed.

The antibackground (ABG) trigger veto was in effect in several early data
sets (see section~\ref{sec:experiment}). ABG energy thresholds in
software trigger varied approximately from -50\% to +50\% to estimate
the influence on the result uncertainty (\showtableref
row \ref{syst:abgthr}).

The calorimeter trigger thresholds were increased by 25\% to estimate
the corresponding contribution to the result uncertainty (\showtableref
 row \ref{syst:trgthr}).

Systematic uncertainties due to RND events usage were estimated by the
fluctuations of the result caused by change of a subset of noise and
background events applied to the simulation (\showtableref
 row \ref{syst:rnd}).

Final-state radiation effect accounted for by the PHOTOS package is
about 4\%. Its systematic uncertainty is estimated to be of about 10\%
of the effect, i.e. $\sim0.4$\%~\cite{photos} (\showtableref
 row \ref{syst:photos}).

The result fluctuated slightly when changing the number of $\theta$ bins to
divide \ee data. The maximal fluctuation for 5, 10, 15, 20 bins
(instead of 4 by default) was taken as a contribution to a systematic
uncertainty (\showtableref row \ref{syst:fit}).

The ToF measurement contribution (\showtableref row
\ref{syst:tofeff}) is described separately in section \ref{sec:tofeff}.

The trigger efficiency was found to be more than 99.3\% for primary
trigger and $98.0\div99.9$\% (depending on the data set) for secondary
trigger. Their variations in error limits give the contribution to
systematic uncertainty (\showtableref row
\ref{syst:PTeff}).

Accuracy of the theoretical formulae used (\showtableref
 row \ref{syst:theory}) is expected to be at the
level of 0.1\%~\cite{KuraevFadinEng}.

\section{Results and conclusion}
Nine data sets recorded by the KEDR detector in the \psip region were
processed, the total number of \psip being about $4\times10^6$.
Our results of the measurement for each data set are listed in
Table~\ref{table:results}.
\begin{table}[h]
\caption{Result and errors (eV) and fit quality for each data set.
}\label{table:results}
\newcommand{\resulthline}{\hline}
\centering\begin{tabular}{|l|c|c|c|c|}\hline
\multicolumn{1}{|c|}{Data set} & \GBmumu & $\sigma_\stat$ &
$\sigma_\syst$ & \rule[-0.5em]{0cm}{1.5em}$\chi^2/\ndf$\\\hline\hline
\Point{1} & 20.5 & 1.2 & 1.2 & 2.6/6\\\resulthline
\Point{2} & 21.5 & 1.7 & 1.3 & 12.6/6\\\resulthline
\Scan{1} & 18.9 & 1.9 & 1.0 & 27.0/26\\\resulthline
\Point{3} & 17.5 & 0.8 & 0.8 & 2.1/6\\\resulthline
\Point{4} & 20.2 & 0.8 & 0.8 & 6.0/6\\\resulthline
\Point{5} & 19.3 & 0.7 & 0.9 & 12.2/6\\\resulthline
\Scan{2} & 20.9 & 1.0 & 1.0 & 28.4/30\\\resulthline
\Scan{3} & 16.1 & 1.3 & 1.4 & 25.0/18\\\resulthline
\Scan{4} & 19.3 & 0.6 & 0.9 & 20.5/18\\\resulthline

\end{tabular}
\end{table}

The final average value is:
\[
\GBmumu = \Result \pm \Stat \pm \Syst ~\eV.
\]

\begin{figure}[ht]
\centering\includegraphics[width=\columnwidth]{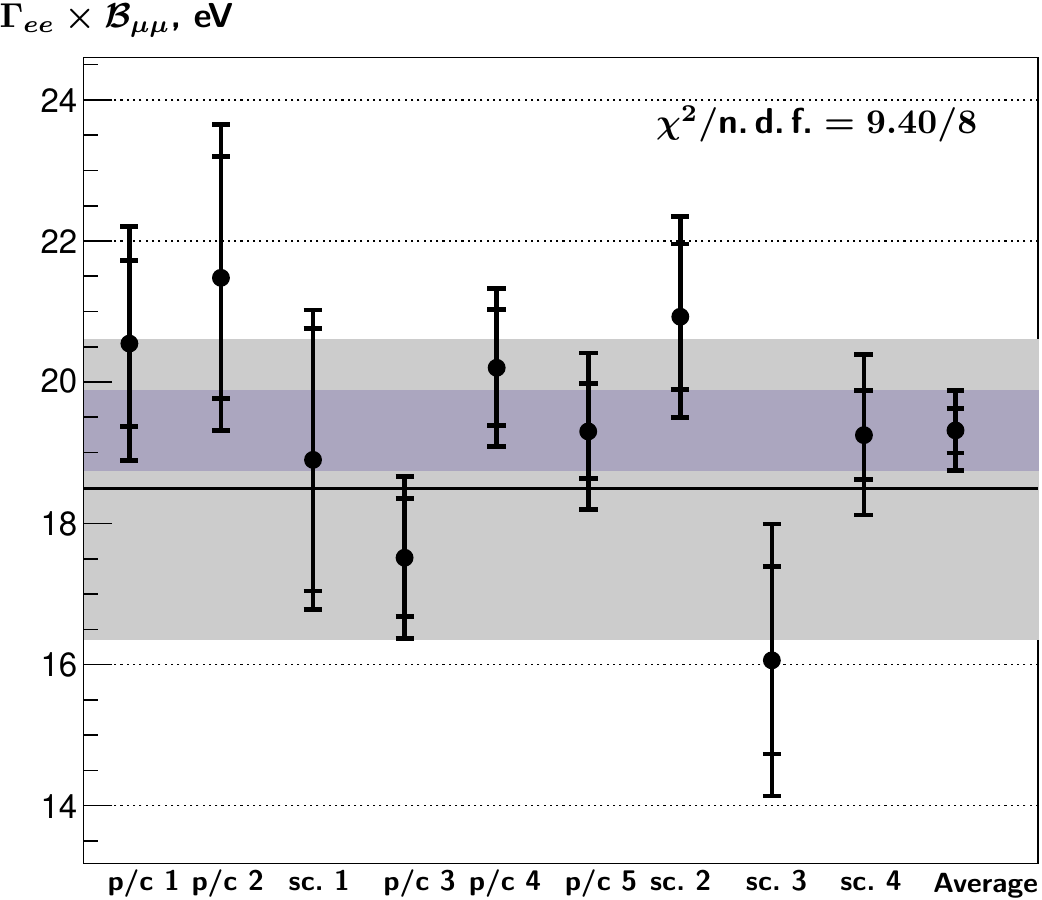}
\caption{Result on $\Gamma_{ee}\times\mathcal{B}_{\mu\mu}$ for each data
set
(ticks on the error bars correspond to statistical and total
uncertainties) and the averaged result with its total error
(the dark gray band). The horizontal line and light gray band
indicate the ``world average'' and its error. The chi-square of the
averaging and number of degree
of freedom are also shown.}\label{fig:result}
\end{figure}

\begin{figure}[ht]
\centering\includegraphics[width=0.95\columnwidth]{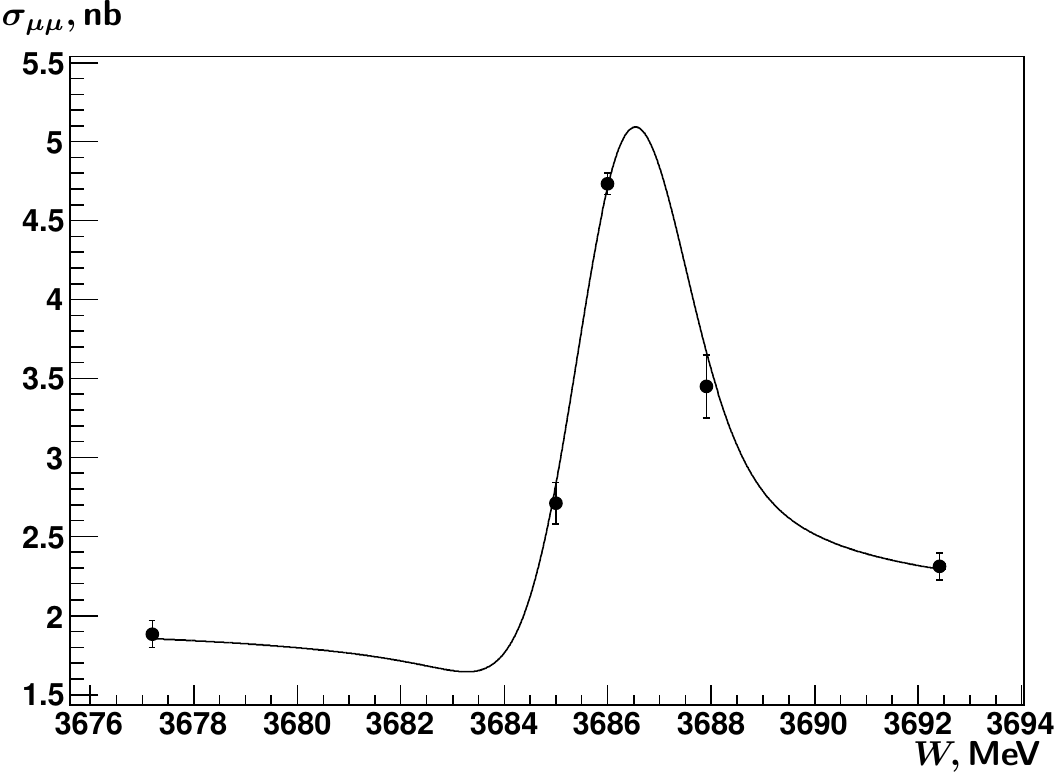}
\caption{An illustration of the observed \mumu
cross section in scan 4.}\label{fig:cs}
\end{figure}

Particle Data Group~\cite{PDG-2016} does not list any direct \GBmumu
measurement. Using
PDG numbers for
$\Gee = \pdgGee \pm \pdgeGee$ keV and
$\Bmumu = (\pdgBmumu \pm \pdgeBmumu)\times 10^{-3}$,
one could get the ``world average'' of
\[
\GBmumu = \pdgResult \pm \pdgError ~\eV,
\]
which is in good agreement with our result (Fig.~\ref{fig:result}).
An example of observed \mumu cross section is shown in Fig.~\ref{fig:cs}.

Combining our \GBmumu
 result with the KEDR measurement of
$\GBhadr = 2.233 \pm 0.015 \pm 0.042 ~\kev$~\cite{KEDR:main_parameters2012}
we obtained the following value of the \psip
electronic width:
\[
    \Gee = \GeeLUResult \pm \GeeLUStat \pm \GeeLUSyst ~\kev
\]
in the assumption of the lepton universality.

Although not presented as a result of this work, the \GBee value also
comes out of the analysis:
\[
\GBee = \eeResult \pm \eeStat \pm \eeSyst ~\eV.
\]
For \psip the resonance
cross section is far too small compared to Bhabha, so the systematic
error in the \ee channel is much bigger. Nevertheless,
this allows us to calculate the \Gee value
without the lepton universality assumption
(the required value of
$\GBtautau = 9.0 \pm 2.6 ~\eV$ is taken from another KEDR
measurement~\cite{KEDR:tau_mass_2007}):
\[
\Gee = \GeeResult \pm \GeeStat \pm \GeeSyst ~\kev.
\]

\begin{figure}[t]
\centering\includegraphics[width=\columnwidth]{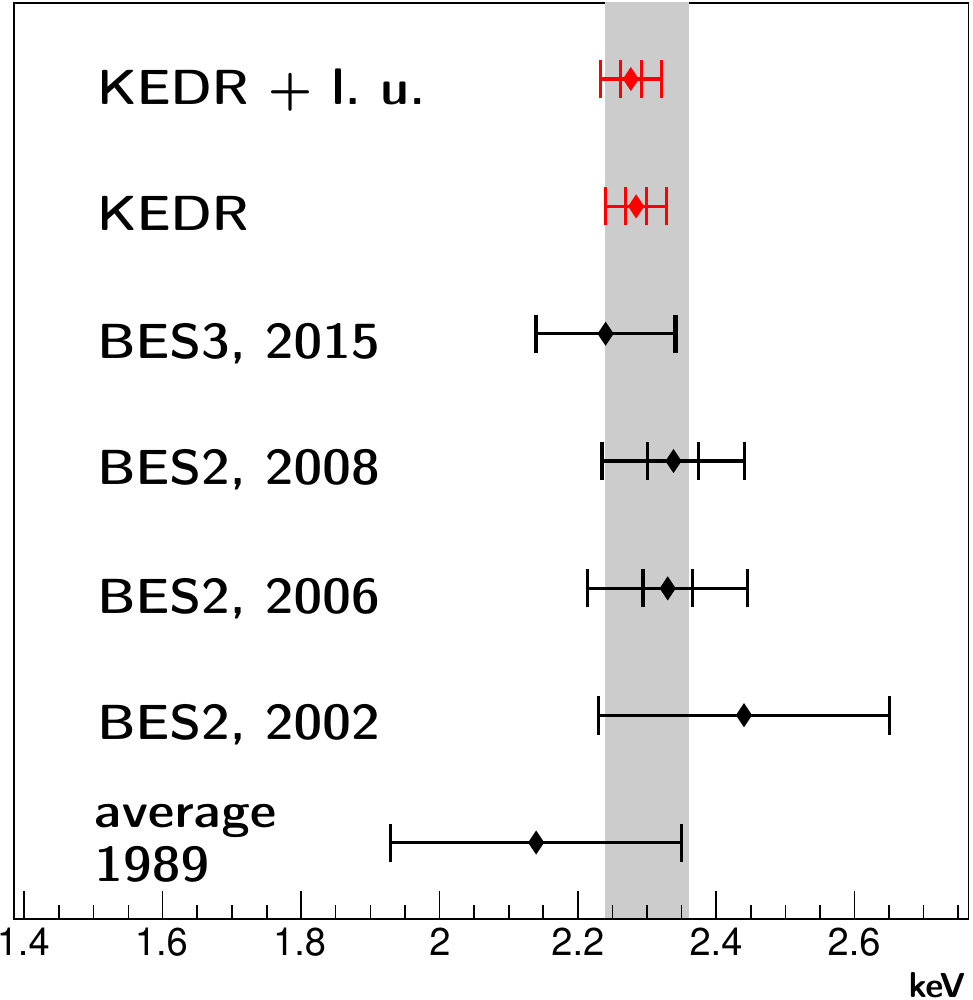}
\caption{Comparison of $\Gee(\psip)$ measurements. The gray band indicates
the current PDG average. Both KEDR values with and without assumption of lepton
universality are represented. Total and statistical (where possible)
errors are shown.}\label{fig:gee}
\end{figure}

Our \Gee values and their comparison with previous results are
presented in the Fig.~\ref{fig:gee}. \Gee uncertainties
are dominated by \GBhadr uncertainties
 either with the lepton universality assumption or without
it.

\section*{Acknowledgments}
We greatly appreciate the efforts of the staff of VEPP-4M to provide
good operation of the complex and the staff of experimental
laboratories for the permanent support in preparing and performing
this experiment.

 The Siberian Branch of the Russian Academy
of Sciences Siberian Supercomputer Center and Novosibirsk State University
Supercomputer Center are gratefully
acknowledged for providing supercomputer facilities~\cite{NSC_SCN}.

This work was supported by Russian Science Foundation under
project N 14-50-00080.

\section*{References}
\bibliography{article}

\end{document}